\documentclass[journal=jpclcd, layout=twocolumn]{achemso}

\usepackage{amsmath,amssymb}

\newcommand{\sign}{\, {\rm sign}\,}

\author{Denys I. Bondar}
\email{dbondar@princeton.edu}
\affiliation{Princeton University, Princeton, NJ 08544, USA}

\author{Renan Cabrera}
\affiliation{Princeton University, Princeton, NJ 08544, USA} 

\author{Andre Campos}
\affiliation{Princeton University, Princeton, NJ 08544, USA} 

\author{Shaul Mukamel}
\affiliation{University of California, Irvine, Irvine, CA 92697, USA} 

\author{Herschel A. Rabitz}
\affiliation{Princeton University, Princeton, NJ 08544, USA} 

\title{Wigner-Lindblad equations for quantum friction} 

\begin{document}

\section{Abstract}
Dissipative forces are ubiquitous and thus constitute an essential part of realistic physical theories. However, quantization of dissipation has remained an open challenge for nearly a century. We construct a quantum counterpart of  classical friction, a velocity-dependent  force acting against the direction of motion. In particular, a translationary invariant Lindblad equation is derived satisfying the appropriate dynamical relations for the  coordinate and momentum (i.e., the Ehrenfest equations). Numerical simulations establish that the model  approximately equilibrates. These findings significantly advance a long search for a universally valid Lindblad model of quantum friction and open opportunities for exploring novel dissipation phenomena.
\begin{figure}
	\includegraphics[width=2in]{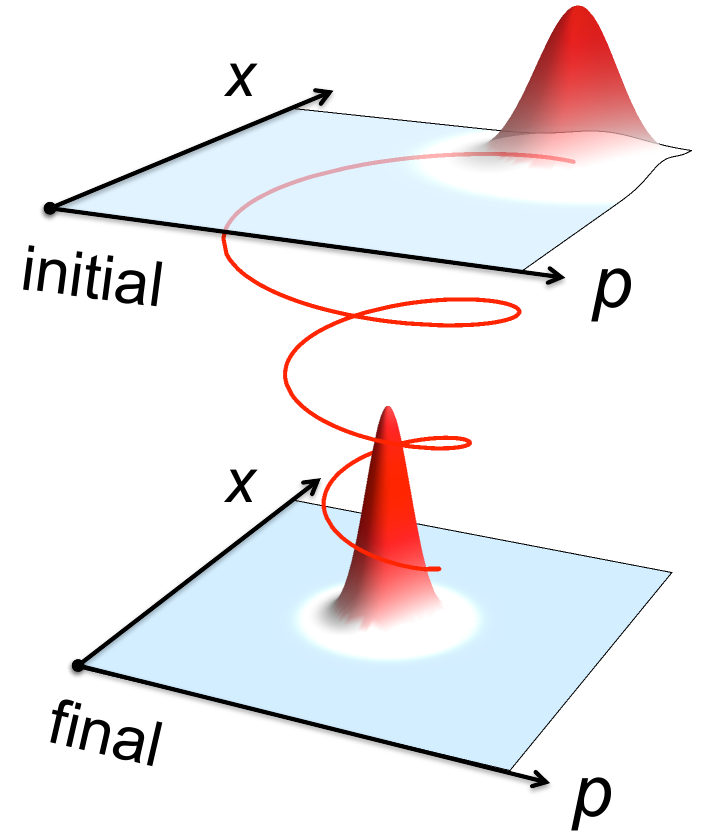}
\end{figure}

\section{Introduction}
Realistic models of quantum systems must include dissipative interactions with an environment, which may be of various nature ranging from a vacuum to a generic thermal bath. Nevertheless, construction of physically consistent quantum models of dissipative forces has been a long standing problem since the birth of quantum mechanics (see, e.g., \cite{Gardiner2004, Razavy2005, Bolivar2012, Caldeira2014}). A common framework for describing open quantum systems is to represent the state of the system by a density matrix, whose evolution is governed by the Lindblad equation \cite{Kossakowski1972, Lindblad1976}. In this Letter, we construct a model of  quantum friction, whose classical counterpart is a velocity-dependent force acting against particle's motion. 

\begin{figure}
		\includegraphics[width=1\hsize]{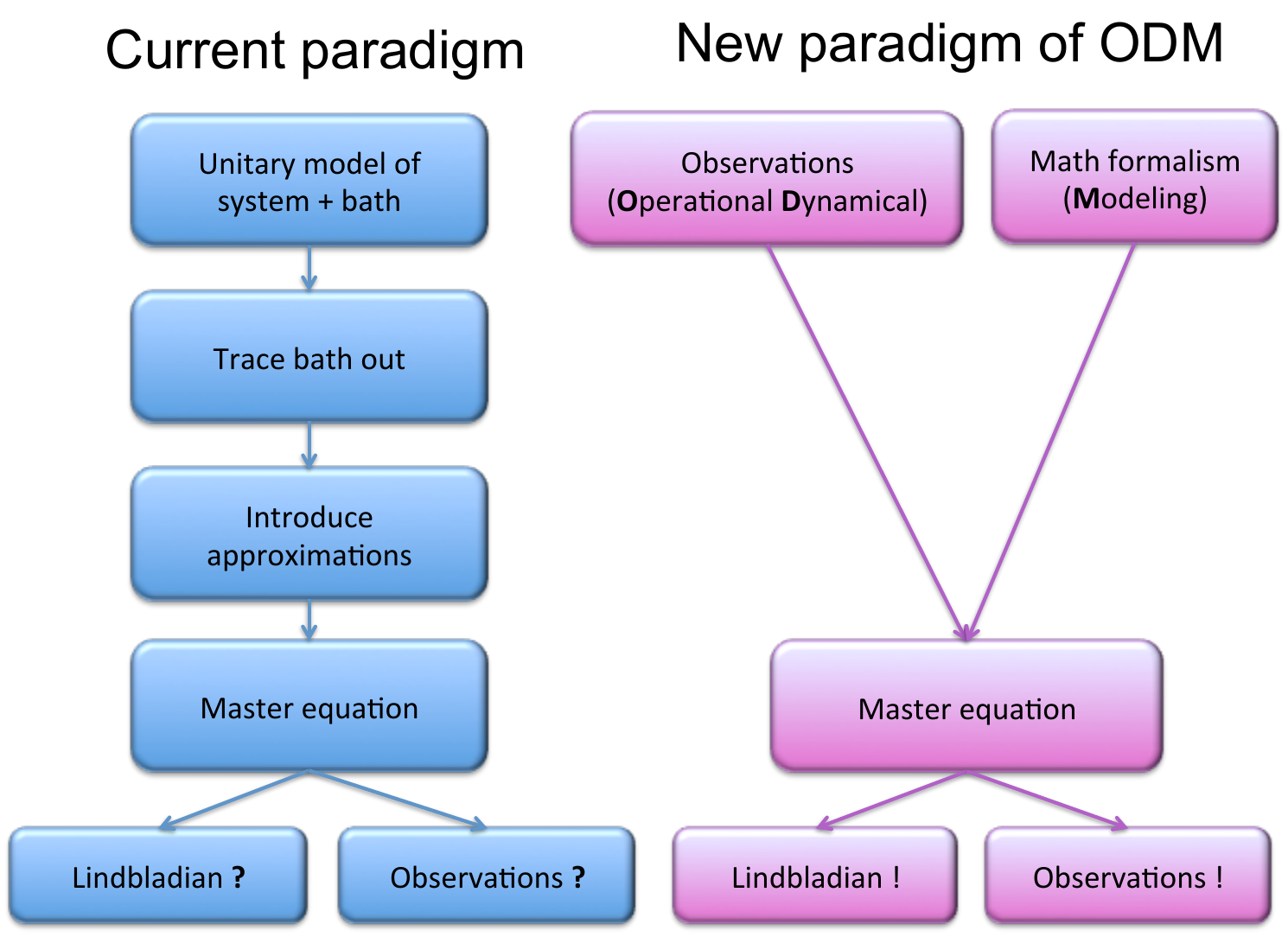}
		\caption{Currently paradigm for deriving master equations governing open system dynamics vs proposed novel approach of Operational Dynamic Modeling (ODM).}
			\label{Fig_ODM_open_system}
\end{figure}
\begin{figure}

		\includegraphics[width=1\hsize]{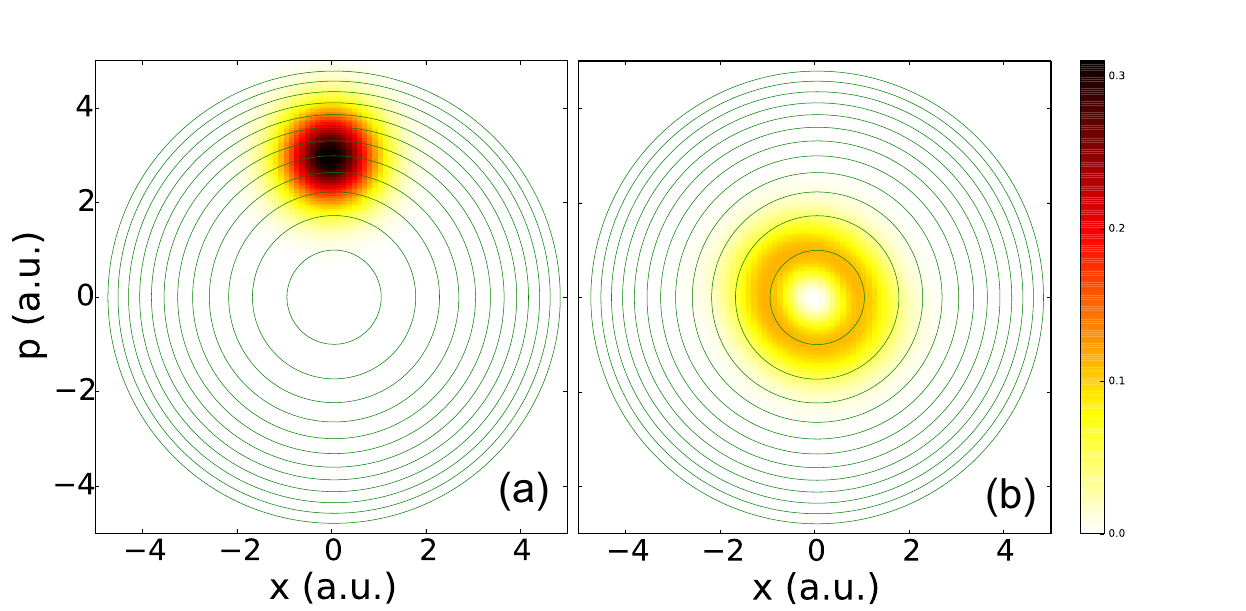}
		\caption{The initial (a) and final (b) Wigner functions for the harmonic oscillator evolving according to the model (\ref{LindbladEq}) governing the Ohmic dissipation [with eq (\ref{SmoothenF})], $\gamma = 0.07$ a.u., $L=3$ a.u., and $\mathfrak{D}=0$]. The circular solid lines depict the level set of the Hamiltonian $H = (p^2+ x^2 )/2$ a.u. (a) The Wigner function of the ground state displaced along the momentum axis. The reached steady state (b) is not a Gaussian distribution.}
			\label{Fig_ZeroTemperatureStedyState}
\end{figure}
\begin{figure}
		\includegraphics[width=1\hsize]{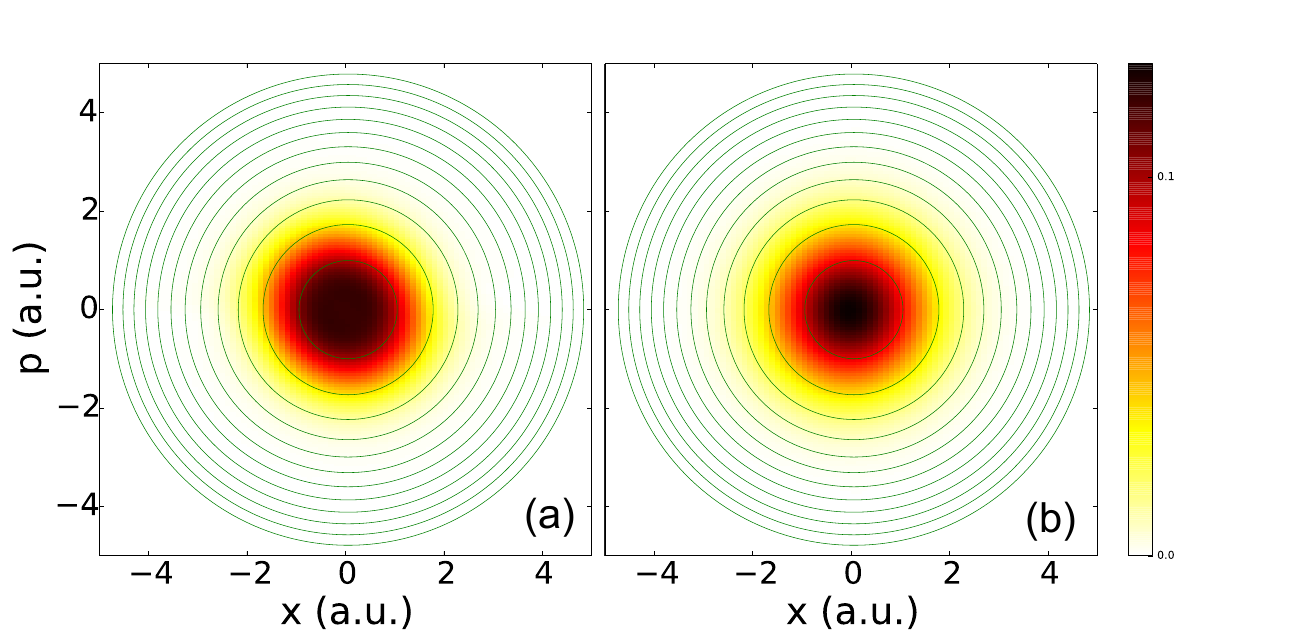}
		\caption{(a) The final Wigner function for the harmonic oscillator evolving according to the model (\ref{LindbladEq}) governing the Ohmic dissipation [with eq (\ref{SmoothenF}), $\gamma = 0.07$ a.u., $L=3$ a.u., and $\mathfrak{D}=0.0143$ a.u.]. The circular solid lines depict the level set of the Hamiltonian $H = (p^2+ x^2 )/2$ a.u. The initial Wigner function is shown in figure \ref{Fig_ZeroTemperatureStedyState}(a). Note that the steady state approaches the thermal Boltzmann state with $kT =  1.166$ a.u. depicted in (b).}
			\label{Fig_StedyStateWithDephasing}
\end{figure}
\begin{figure}
		\includegraphics[width=1\hsize]{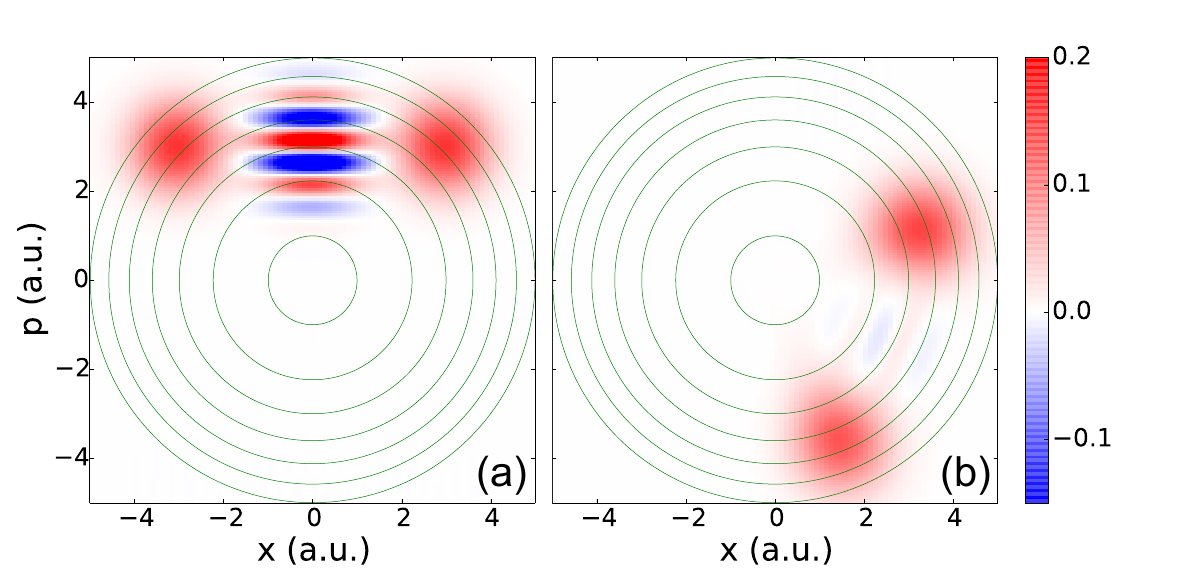}
		\caption{(a) The Wigner function of the Schr\"{o}dinger cat state at time $t=0$. (b) The Wigner function at later time $t=2$ a.u. after evolving according to the model (\ref{LindbladEq}) governing the Ohmic dissipation (system's parameters are defined in figure \ref{Fig_StedyStateWithDephasing}.). As time progresses, the Wigner function's negativity vanishes and the state approaches the Boltzmann equilibrium shown in figure \ref{Fig_StedyStateWithDephasing}(b).}
			\label{Fig_SchrodingerCatDissipative}
\end{figure}
\begin{figure}
		\includegraphics[width=1\hsize]{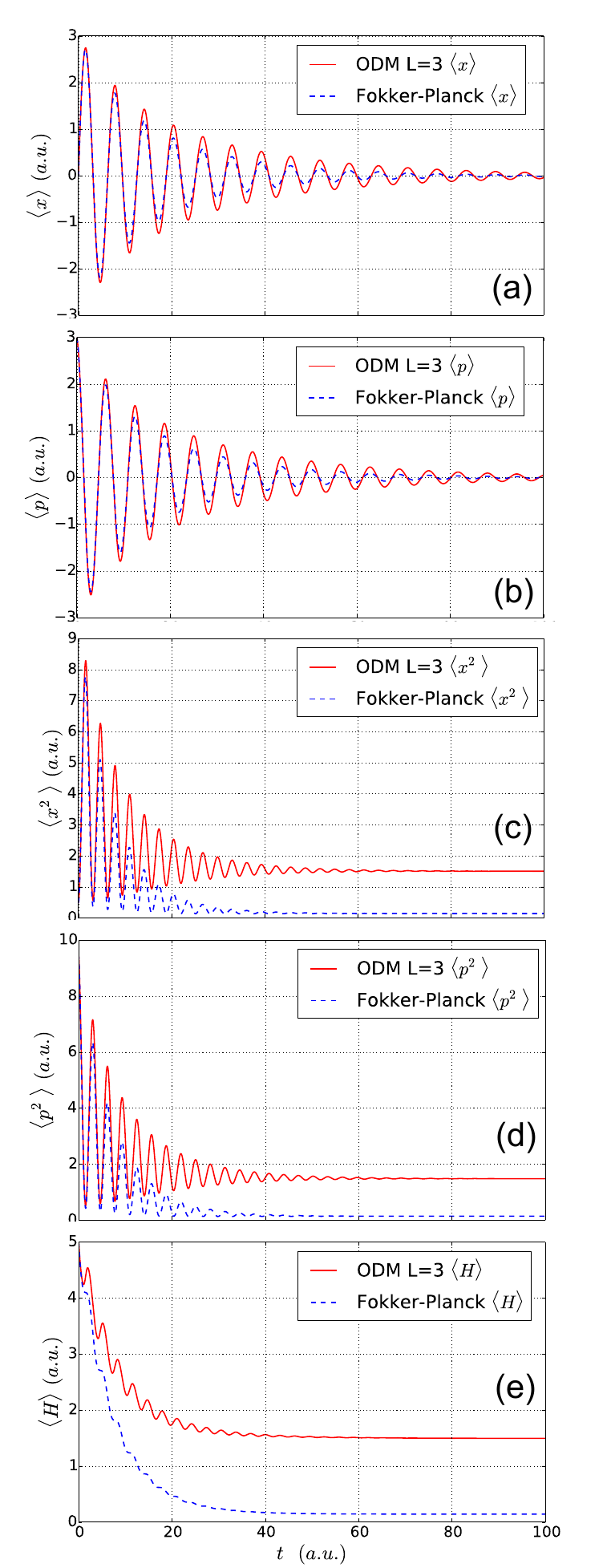}
		\caption{Quantum (solid red lines) [eq (\ref{LindbladEq})] vs classical (dashed blue lines) [eq  (\ref{ClassicalLimit})] dissipative dynamic of a harmonic oscillator. Parameters for both systems are identical (parameters are specified in figure \ref{Fig_StedyStateWithDephasing}.) Time-evolution of the first [(a), (b)] and second [(c), (d)] order moments. (e) The total energy variation.}
			\label{Fig_ODMvsFokkerPlanck}
\end{figure}

By employing the phase space representation of quantum mechanics  \cite{Curtright1998, Curtright2011, Curtright2013}, where an observable $O=O(x,p)$ is assumed to be a real-valued function of coordinate $x$ and momentum $p$, and system's state is represented by the Wigner function $W=W(x,p)$, we derive the Lindblad-Wigner equation 
\begin{align}
	\frac{d}{dt} W =& -\frac{i}{\hbar} (H \star W - W \star H ) \notag\\
	& \qquad + D[W] + D'[W], \label{LindbladEq} \\
	H =& p^2 / (2m) + U(x), \label{HamiltonianDef}\\
	D[W] =& \frac{2\gamma}{\hbar} \Big( A \star W \star A^*  - \frac 12 W \star A^* \star A \notag\\
	&  - \frac 12 A^* \star A \star W \Big), \label{DeffDissipator} \\
	D'[W] =& \frac{2\mathfrak{D}}{\hbar^2}\Big(  x \star W \star x  
	 - \frac 12 W \star x \star x  \notag\\
	 &- \frac 12 x \star x \star W \Big) = \mathfrak{D} \frac{\partial^2 W}{\partial p^2}, \label{DephasingDiss}\\
	\star = &\exp \frac{i\hbar}{2} \left( 
		\overleftarrow{\frac{\partial}{\partial x}} \overrightarrow{\frac{\partial}{\partial p}} -
		\overleftarrow{\frac{\partial}{\partial p}} \overrightarrow{\frac{\partial}{\partial x}}
	 \right), \label{MoyalStar}
\end{align}
which guarantees completely positive dynamics of the density matrix underlying the Wigner function $W$ for an arbitrary operator $A$. In standard derivations, one finds a family of  relaxation  operators  $A$ by  assuming a  weak coupling to a bath and expanding the dynamics perturbatively. Here we adopt a different strategy: We require that the first moments of $W$ satisfy the Ehrenfest  equations
\begin{align}
	&\frac{d}{dt} \langle x \rangle 		=  \frac{1}{m} \langle p \rangle,	 
			\label{Ehrenfest1} \\
	&\frac{d}{dt} \langle p \rangle		= -\langle U'(x) \rangle - 2\gamma \langle \sign(p)f(|p|) \rangle,
			 \label{Ehrenfest2} 
\end{align}
characterizing motion of a particle of mass $m$ interacting with an environment induced velocity-dependent friction. The conventional derivations of master equations (see figure \ref{Fig_ODM_open_system})  do not guaranteed satisfaction of these relations. Using  the  Operational Dynamical Modeling (ODM) algorithm to be described below,  we construct an operator $A$ that satisfies the constraints (\ref{Ehrenfest1}) and (\ref{Ehrenfest2})
\begin{align}
	A =& \sqrt{ L f\left(|p| + \frac{\hbar}{2L} \right) } \exp(-i\sign(p)x/L). \label{FinalA} 
\end{align}
The classical limit  of the Lindblad-Wigner equation (\ref{LindbladEq}) with eq (\ref{FinalA})  recovers the appropriate Fokker-Planck equation  \cite{Gardiner1985},
\begin{align}\label{ClassicalLimit}
	D[W] =  2 \gamma  \frac{\partial}{\partial p} \left[ \sign(p) f(|p|) W \right] + O \left( \hbar \right).
\end{align} 
The Ehrenfest relations for the second moments may also be obtained from eq (\ref{LindbladEq}):
\begin{align}
	\frac{d}{dt} \langle p^2 \rangle		&= -2\langle pU'(x)  \rangle  \notag\\
	& - 2 \gamma \left\langle f(|p|)\left(2|p|   - \frac{\hbar}{ L} \right)  \right \rangle + 2\mathfrak{D}, 
			\label{EhrenfestNew3} \\
	\frac{d}{dt} \langle xp \rangle	&= \frac{1}{m} \langle p^2 \rangle - \langle xU'(x) \rangle \notag\\
		& \qquad - 2\gamma \langle \sign(p) f(|p|)x  \rangle, \label{EhrenfestNew4} \\
	 \frac{d}{dt} \langle x^2 \rangle		&= \frac{2}{m} \langle xp \rangle + \frac{\gamma\hbar L}{2}  \left\langle \frac{f'(|p|)^2}{f(|p|)} \right\rangle.
			\label{EhrenfestNew5}
\end{align}
In order to employ eq (\ref{LindbladEq}), the following free parameters must be specified i) $f(p) \geq 0$ -- the velocity dependence of the dissipative force, ii) $\gamma \geq 0$ -- a friction coefficient, iii) $L>0$ -- a length-scale constant defining the dynamics of second-order moments (\ref{EhrenfestNew3})-(\ref{EhrenfestNew5}), and iv) $\mathfrak{D} \geq 0$ is a dephasing constant chosen such that dynamics equilibrates to the Boltzmann state with some temperature.

Equation (\ref{MoyalStar}) defines the Moyal product \cite{Curtright1998, Curtright2011, Curtright2013}, which is a result of mapping the non-commutative matrix product in the Hilbert space into the phase space. As a result, the dissipator (\ref{DeffDissipator}) is obtained by Wigner transforming the Lindblad equation for the density matrix, thus the Wigner function's marginals stay positive throughout the entire evolution. The dissipator (\ref{DephasingDiss}) describes dephasing, the loss of quantum coherence \cite{Zurek2003, Gardiner2004, Cabrera2015}; whereas, the dissipator (\ref{DeffDissipator}) causes amplitude damping. The usage of sign and modulus functions in eqs (\ref{FinalA}) and (\ref{Ehrenfest2}) are necessary to ensure that the friction force acts against the particle's motion. The dissipator (\ref{DeffDissipator}) is translationally invariant (more precisely, Galilei-covariant): a spatial displacement $W(x,p) \to W(x+c, p)$ implies $D[W](x,p) \to D[W](x+c, p)$. Numerical simulations establish that the long-time dynamics govern by eq (\ref{LindbladEq}) rigorously does not equilibrate; however, the dynamics can be said to approximately equilibrate. Namely, one can find a value of the dephasing constant $\mathfrak{D}$ in eq (\ref{DephasingDiss}) such that the steady state closely resembles the Boltzmann state with temperature $T$. In this sense, we numerically define the temperature dependence of $\mathfrak{D} = \mathfrak{D}(T)$.

\section{Comparison with other theories}

Current quantum friction models can be roughly divided into the two categories:  i)  Lindblad models not obeying the Ehrenfest relations have been proposed in \cite{Berman1992, Gao1997, Vacchini2009}. The fundamental reason for the Ehrenfest equation violation is the ubiquitous usage of $A$ [eq (\ref{DeffDissipator})], which is taken to be linear with respect to the coordinate and momentum [see the comment after eq (\ref{EqsForAFunctions})].  ii) Non-Lindblad models \cite{Caldeira1983, Ford1987, Ford1988, Hu1992, Bolivar2005, Cleary2011, Bolivar2012, Bolivar2015} obeying the Ehrenfest relations that preserve state's positivity for sufficiently high temperatures. Contrary to the claims, the master equations in \cite{Diosi1993, Vacchini2000, Petruccione2002} belong to the same category. In particular, the model in \cite{Diosi1993} produces uncontrollable heating \cite{Munro1996, Stenholm1997} greatly spreading the wave packet. This state of the field is unsatisfactory because non-Lindblad master equations are known to lead to negative probabilities \cite{Munro1996, Stenholm1997}, whereas the violation of the Ehrenfest equations lead to unphysical artifacts \cite{Wiseman1998}. 
 
A comparative review \cite{Kohen1997} of major quantum dissipation theories further revealed that no existing model is simultaneously i) complete positive, ii) translationally invariant, and iii) asymptotically approaching thermal equilibrium. The present model (\ref{LindbladEq}) exactly obeys the first two properties, whereas the latter can be satisfied approximately (this can be achieved exactly in the free particle case). Furthermore, our simulations confirm that the dynamics of our model does not cause uncontrollable spreading of the wave packet even at zero temperature, and approaches thermal equilibrium at higher temperatures, thereby overcoming computational and physical inconsistencies plaguing  other dissipative theories. The model (\ref{LindbladEq}) is obtained as a unique consequence of the Ehrenfest constrains (\ref{Ehrenfest1}) and (\ref{Ehrenfest2}) and the requirements for the dynamics to be Lindblad, translationary invariant, and  state independent [i.e., $A=A(x,p)$ in eq (\ref{DeffDissipator}) does not depend on the Wigner function]. 
 
 A difficulty of constructing physical models of quantized friction lies in fundamental limitations of the current paradigm for modeling open system dynamics (see figure \ref{Fig_ODM_open_system}). First, the combined system plus bath are assumed to evolve unitarily; second, the environmental degrees of freedom are traced out by making a number of approximations. This procedure neither guarantees that the resultant master equation can reproduce the observations characterizing phenomenon of interest, nor that the equations to have a desired mathematical structure. 

It is noteworthy that the limitations of the current paradigm persist even if no approximation is necessary to trace the bath out. For example, the  Hu-Paz-Zhang master equation \cite{Ford1987, Ford1988, Hu1992} for a harmonic oscillator  interacting  with  a  linear  passive  heat  bath  of  oscillators  is exact under the assumption that the bath is initially at equilibrium and not coupled to the oscillator. The obtained non-Lindblad master equation preserves the density matrix's positivity and satisfies the Ehrenfest relations. However, the Hu-Paz-Zhang propagation of  states initially uncorrelated with the environment leads to instantaneous infinite spreading of the wave packet \cite{Ford2001}, which could be fixed by modulating the friction coefficient during the evolution \cite{Fleming2011}. Further revealed problems have lead to the conclusion  that the model is of very limited physical utility \cite{Ford2005}. 

To overcome these fundamental limitations, a new paradigm of ODM \cite{Bondar2011c} has been recently put forth, enabling the generation of models directly from observed data (see figure \ref{Fig_ODM_open_system}). To derive master equations, ODM needs two inputs: observed data recast in the form of Ehrenfest relations and a specified mathematical structure of the equation of motion. As an outcome, ODM guarantees that the resulting equations of motion have the desired physical structure to reproduce the supplied dynamical observations. This formalism has provided new interpretation of the Wigner function \cite{Bondar2013a}, unveiled conceptual inconstancies in finite-dimensional quantum mechanics \cite{Bondar2011}, formulated dynamical models in topologically nontrivial spaces \cite{Zhdanov2015}, advanced the study of quantum-classical hybrids \cite{Radonjic2014}, and lead to development of efficient numerical techniques \cite{Cabrera2015, Cabrera2014}.

\section{Derivation}

We begin by identifying the amplitude dumping dissipator $D[W]$ [eq (\ref{DeffDissipator})], thus  the dephasing coefficient $\mathfrak{D}$ is set to zero to ignore $D'[W]$ [eq (\ref{DephasingDiss})]. Substituting eq (\ref{LindbladEq}) into eqs (\ref{Ehrenfest1}) and (\ref{Ehrenfest2}) and then dropping the averaging, which is justified by $A$ being state independent, we obtain equations for an unknown function $A = A(x,p)$, defining the Lindblad dissipator (\ref{DeffDissipator}),
\begin{align}
	A^* \star \frac{\partial A}{\partial p} - \frac{\partial A^*}{\partial p} \star A &= 0, \\
	A^* \star \frac{\partial A}{\partial x} - \frac{\partial A^*}{\partial x} \star A &= -4i\sign(p)f(|p|).  					\label{EqsForAFunctions}
\end{align}

The Lindblad models for Omic friction, $f(p) = |p|$, has  been widely studied (see, e.g., \cite{Lindblad1976a, Berman1992, Diosi1993, Gao1997, Vacchini2000, Ferialdi2016}), where $A$ was found to be a linear combination of $x$ and $p$. However, as we shall now  establish, \emph{no Lindblad dynamics with an $A$ linear in $x$ and $p$ satisfies the Ehrenfest equations (\ref{Ehrenfest1}) and (\ref{Ehrenfest2})}. Indeed, substituting $A = a x + b p$ into eq (\ref{EqsForAFunctions}) leads to
\begin{align}
	( a^* b - a b^*) x = 0, \quad ( a^* b - a b^*) p = 4ip, 
\end{align}
where a contradiction becomes evident. This conclusion holds in the case of Lindblad models with several such $A$ operators. Our model (\ref{FinalA}) circumvents this no-go result due to its new non-linear dependence on $x$ and $p$. 

The action of the dissipative force is expected to be translationary invariant.  One observes directly from the definition of the Moyal product (\ref{MoyalStar}) that if 
\begin{align}\label{TransInvarLemma1}
	A(x,p) = g(p) \exp(iCx), \quad C^* = C,
\end{align}
then the dissipator $D[W]$ (\ref{DeffDissipator}) is translationary invariant. Formally, the dissipator $D[W]$ with $A$ given by eq (\ref{TransInvarLemma1}) obeys
\begin{align}\label{TransInvarLemma2}
	D[W] = F\left( p, \frac{\partial}{\partial x}, \frac{\partial}{\partial p}, \frac{\partial^2}{\partial x^2}, \frac{\partial^2}{\partial x\partial p}, \frac{\partial^2}{\partial p^2}, \cdots \right) W,
\end{align}
where $\cdots$ denotes higher order derivatives, and the function $F$ explicitly  does not depend on $x$. Additionally, eq (\ref{Ehrenfest1}) is satisfied  for any real valued $g(p)$. Therefore, substituting the expansion
\begin{align}\label{PerturbativeExpansionG}  
	g(p) = \sum_{n=0}^{\infty} g_n(p) \hbar^n
\end{align}
into eqs (\ref{TransInvarLemma1}) and (\ref{EqsForAFunctions}), we recursively finding  all terms in expansion (\ref{PerturbativeExpansionG}) (see the supporting information). 
Finally, the resultant series can be then summed up, thereby leading to the exact solution (\ref{FinalA}) where $L$ is a positive length-scale constant to balance the dimensionality. The value of $L$ dictates the time-dynamics of second-order moments [eqs (\ref{EhrenfestNew3})-(\ref{EhrenfestNew5})].

A simple steady state solution is found by additionally requiring the spatial homogeneity
\begin{align}
	& \frac{\partial W}{\partial x} = 0, \quad \frac{d W}{d t} = 0, \quad U =0, \quad \mathfrak{D} = 0 \notag\\
	& \Longrightarrow W = \mbox{const} / f(|p|).
\end{align}
Therefore, if $f(p)$ is chosen to be an inversely proportional to a thermal equilibrium state, the free particle dynamics of the model (\ref{FinalA}) equilibrates without the need for the dephasing dissipator $D'[W]$ (\ref{DephasingDiss}).

Both quantum corrections in eqs (\ref{EhrenfestNew3}) and  (\ref{EhrenfestNew5}), i.e., the terms proportional to $\hbar$, are position independent, thereby reconfirming the translational invariance. According to eq (\ref{EhrenfestNew5}), the quantum correction to Ohmic dissipation (with $f(p) = |p|$) is singular, therefore the function $f(p)$ should be regularized. For example, we employ the following smoothing in numerical simulations  (see, e.g., figures \ref{Fig_ZeroTemperatureStedyState}-\ref{Fig_ODMvsFokkerPlanck}) 
\begin{align}\label{SmoothenF}
	f(p) = p^2 / \sqrt{ p^2 + \epsilon^2 }, \qquad \epsilon^2 = 0.5.
\end{align}  
The Ehrenfest relations (\ref{Ehrenfest1}), (\ref{Ehrenfest2}), (\ref{EhrenfestNew3})-(\ref{EhrenfestNew5}) have been verified numerically for diferent values of parameters and initial conditions. 

Equation (\ref{EhrenfestNew5}) establishes that the steady state need not  coincide with thermal equilibrium. As an example, consider a harmonic oscillator. The equilibrium state is characterized by the identity $\langle xp \rangle = 0$, which contradicts the steady state condition $d \langle x^2 \rangle / dt = 0$ because the quantum correction in eq (\ref{EhrenfestNew5}) is strictly positive. Furthermore, if the steady state is positive, then its Wigner function should be more pronounced in the second and fourth quadrants of the phase space (where $xp < 0$) to compensate for the quantum correction [this small asymmetry as can be noticed in figures \ref{Fig_ZeroTemperatureStedyState}(b) and \ref{Fig_StedyStateWithDephasing}(a)]. 

Figure \ref{Fig_ZeroTemperatureStedyState} shows the initial state with average momentum $p=3$ a.u. (arbitrary units, $\hbar=m=1$, are employed in the simulations) reaches the steady state with a circularly shaped  Wigner function. The latter has a characteristic asymmetry, as discussed above. The reached state does not resemble the Boltzmann thermal equilibrium.

To allow the dynamics to equilibrate approximately, we include the dephasing dissipator $D'[W]$ [eq (\ref{DephasingDiss})] with a non-vanishing $\mathfrak{D}$. In the case of a harmonic oscillator [$U(x) =m\omega^2 x^2 / 2$], the fluctuation-dissipation theorem follows from the second order Ehrenfest relations (\ref{EhrenfestNew3})-(\ref{EhrenfestNew5}):
\begin{align}
	\frac{\mathfrak{D}}{\gamma} =& 
	\left\langle f(|p|)\left(2|p|   - \frac{\hbar}{ L} \right)  \right. \notag\\
	& \left.  \qquad -  \hbar L \left(\frac{ m\omega}{2}\right)^2 \frac{f'(|p|)^2}{f(|p|)} \right\rangle_{st},
\end{align}
where $\langle \cdots \rangle_{st}$ denotes averaging over a steady state.

The steady state for the model with $\mathfrak{D}=0.0143$ (a.u.) is shown in figure \ref{Fig_StedyStateWithDephasing}. For a sufficiently high value of $\mathfrak{D}$, the ring in figure \ref{Fig_ZeroTemperatureStedyState}(b) is washed out and the Wigner function of the steady state looks like a gaussian [figure \ref{Fig_StedyStateWithDephasing}(a)], which well approximates the Boltzmann equilibrium for some temperature [figure \ref{Fig_StedyStateWithDephasing}(b)]. The larger the dephasing coefficient $\mathfrak{D}$, the more accurate the equilibration dynamics. Additionally, we have also verified that the approximate equilibration dynamics occurs in the case of anharmonic oscillators. Figure \ref{Fig_SchrodingerCatDissipative} establishes that the evolution generated by the model (\ref{LindbladEq}) washes out the quantum interference initially present in the Schr\"{o}dinger cat state, while the dynamics equilibrates.

Despite a simple look of the Wigner functions in figures \ref{Fig_ZeroTemperatureStedyState} and \ref{Fig_StedyStateWithDephasing}, full  time-dependent dynamics are rich in quantum features. Figure \ref{Fig_ODMvsFokkerPlanck} compares quantum dissipative dynamics, governed by the Lindblad-Wigner equation (\ref{LindbladEq}), with the corresponding classical Fokker-Planck evolution (\ref{ClassicalLimit}). Even though both quantum and classical master equations satisfy the same first order Ehrenfest theorems (\ref{Ehrenfest1})-(\ref{Ehrenfest2}), time evolution of the expectation values of the coordinate [figure \ref{Fig_ODMvsFokkerPlanck}(a)] and momentum [figure \ref{Fig_ODMvsFokkerPlanck}(b)] exhibit quantitative differences. Since the optical polarizability is proportional to $\langle x \rangle$, the predicted quantum corrections may be observed via non-linear spectroscopy \cite{Mukamel1999}. The  correction to the second order Ehrenfest theorems (\ref{EhrenfestNew3})-(\ref{EhrenfestNew5}), enforcing the Heisenberg uncertainty priciple, qualitative change open system dynamics [figures \ref{Fig_ODMvsFokkerPlanck}(c) and \ref{Fig_ODMvsFokkerPlanck}(d)]. As a result, the expectation value of energy in classical dissipative dynamics monotonically decreases; whereas, energy revives in quantum case at short time scales [figure \ref{Fig_ODMvsFokkerPlanck}(e)].

\section{Outlook} 

In order to describe quantum dissipative dynamics, emerging in many areas of physics, there is a need for a Lindblad model satisfying the Ehrenfest relations (\ref{Ehrenfest1}) and (\ref{Ehrenfest2}), with long-time  dynamics converging to an equilibrium state.  Currently, the lack of such model has been substituted by a  multitude of dissipative theories. Using ODM (figure \ref{Fig_ODM_open_system}), we have found the translationally invariant Wigner-Lindblad model (\ref{LindbladEq}) exactly obeying the Ehrenfest equations (\ref{Ehrenfest1}) and (\ref{Ehrenfest2}). Furthermore, according to numerical simulations, our model not only shows that a state with non-vanishing mean velocity [figure \ref{Fig_ZeroTemperatureStedyState}(a)] approximately approaches the Boltzmann equilibrium (figure \ref{Fig_StedyStateWithDephasing}), but also exhibits pronounce quantum corrections (figure \ref{Fig_ODMvsFokkerPlanck}) even in the case of a harmonic oscillator.

The following Ehrenfest relation ubiquitously arises in molecular dynamics \cite{Mukamel1999}
\begin{align}
	\frac{d}{dt} \langle p \rangle(t) 
		= - m \omega^2 \langle x \rangle(t) - \int_{-\infty}^t d\tau \gamma(t-\tau) \langle p \rangle(\tau),
\end{align}
where the time dependent dissipation coefficient $\gamma(t)$ is connected with the spectral density of the bath, which characterizes the nature of dissipative dynamics. Such a generalization of the developed model requires the application of ODM to non-Markovian dynamics. In this case, the Lindblad-Wigner equation (\ref{LindbladEq}) will have to be replaced by a corresponding time-convolutionless master equation (see, e.g., \cite{Breuer2007}), thus leading to a time dependent extension of the relaxation operator (\ref{FinalA}).

The presented derivation of the master equation directly from time evolution of expectation values embodied in Ehrenfest relations, is a long-sought alternative to the current cumbersome paradigm for obtaining equations of motions (see figure \ref{Fig_ODM_open_system}). A master equation is typically obtained by performing a number of approximations after the bath is traced out of a combined system-bath model. Such a derivation usually leads to either a non-Lindblad master equation or a model incapable of reproducing observations. The presented ODM-based derivation overcomes all these fundamental weaknesses by deriving Lindblad equations enforced to be compatible with the Ehrenfest equations. This formalism opens new horizons in quantum non-equilibrium statistical mechanics. 

\begin{acknowledgement}
D.I.B., R.C., H.A.R. respectively acknowledge financial support from NSF CHE 1058644, DOE DE-FG02-02-ER-15344 and ARO-MURI W911NF-11-1-0268. D.I.B. was also supported by 2016 AFOSR Young Investigator Research Program. A.G.C was supported by the Fulbright foundation. S. M.  gratefully acknowledges the support of NSF CHE-1361516 and the Chemical Sciences, Geosciences, and Biosciences division, Office of Basic Energy Sciences, Office of Science, U.S. Department of Energy. We want to thank David Tannor for drawing our attention to \cite{Kohen1997} and Dmitry Zhdanov for numerous insightful discussions.
\end{acknowledgement}

{\bf Supporting Information Available:} Maple code for symbolic derivation of the main results of the paper.

\bibliography{quatum_brownian}

\providecommand{\latin}[1]{#1}
\providecommand*\mcitethebibliography{\thebibliography}
\csname @ifundefined\endcsname{endmcitethebibliography}
  {\let\endmcitethebibliography\endthebibliography}{}
\begin{mcitethebibliography}{43}
\providecommand*\natexlab[1]{#1}
\providecommand*\mciteSetBstSublistMode[1]{}
\providecommand*\mciteSetBstMaxWidthForm[2]{}
\providecommand*\mciteBstWouldAddEndPuncttrue
  {\def\EndOfBibitem{\unskip.}}
\providecommand*\mciteBstWouldAddEndPunctfalse
  {\let\EndOfBibitem\relax}
\providecommand*\mciteSetBstMidEndSepPunct[3]{}
\providecommand*\mciteSetBstSublistLabelBeginEnd[3]{}
\providecommand*\EndOfBibitem{}
\mciteSetBstSublistMode{f}
\mciteSetBstMaxWidthForm{subitem}{(\alph{mcitesubitemcount})}
\mciteSetBstSublistLabelBeginEnd
  {\mcitemaxwidthsubitemform\space}
  {\relax}
  {\relax}

\bibitem[Gardiner and Zoller(2004)Gardiner, and Zoller]{Gardiner2004}
Gardiner,~C.; Zoller,~P. \emph{Quantum noise: a handbook of Markovian and
  non-Markovian quantum stochastic methods with applications to quantum
  optics}; Springer, 2004\relax
\mciteBstWouldAddEndPuncttrue
\mciteSetBstMidEndSepPunct{\mcitedefaultmidpunct}
{\mcitedefaultendpunct}{\mcitedefaultseppunct}\relax
\EndOfBibitem
\bibitem[Razavy(2005)]{Razavy2005}
Razavy,~M. \emph{Classical and quantum dissipative systems}; World Scientific,
  2005\relax
\mciteBstWouldAddEndPuncttrue
\mciteSetBstMidEndSepPunct{\mcitedefaultmidpunct}
{\mcitedefaultendpunct}{\mcitedefaultseppunct}\relax
\EndOfBibitem
\bibitem[Bolivar(2012)]{Bolivar2012}
Bolivar,~A.~O. The dynamical-quantization approach to open quantum systems.
  \emph{Annals Phys.} \textbf{2012}, \emph{327}, 705--732\relax
\mciteBstWouldAddEndPuncttrue
\mciteSetBstMidEndSepPunct{\mcitedefaultmidpunct}
{\mcitedefaultendpunct}{\mcitedefaultseppunct}\relax
\EndOfBibitem
\bibitem[Caldeira(2014)]{Caldeira2014}
Caldeira,~A.~O. \emph{An Introduction to Macroscopic Quantum Phenomena and
  Quantum Dissipation}; Cambridge University Press, 2014\relax
\mciteBstWouldAddEndPuncttrue
\mciteSetBstMidEndSepPunct{\mcitedefaultmidpunct}
{\mcitedefaultendpunct}{\mcitedefaultseppunct}\relax
\EndOfBibitem
\bibitem[Kossakowski(1972)]{Kossakowski1972}
Kossakowski,~A. On quantum statistical mechanics of non-Hamiltonian systems.
  \emph{Rep. Math. Phys.} \textbf{1972}, \emph{3}, 247 -- 274\relax
\mciteBstWouldAddEndPuncttrue
\mciteSetBstMidEndSepPunct{\mcitedefaultmidpunct}
{\mcitedefaultendpunct}{\mcitedefaultseppunct}\relax
\EndOfBibitem
\bibitem[Lindblad(1976)]{Lindblad1976}
Lindblad,~G. On the generators of quantum dynamical semigroups. \emph{Commun.
  Math. Phys.} \textbf{1976}, \emph{48}, 119--130\relax
\mciteBstWouldAddEndPuncttrue
\mciteSetBstMidEndSepPunct{\mcitedefaultmidpunct}
{\mcitedefaultendpunct}{\mcitedefaultseppunct}\relax
\EndOfBibitem
\bibitem[Curtright \latin{et~al.}(1998)Curtright, Fairlie, and
  Zachos]{Curtright1998}
Curtright,~T.; Fairlie,~D.; Zachos,~C. {Features of time-independent Wigner
  functions}. \emph{Phys. Rev. D} \textbf{1998}, \emph{58}, 025002\relax
\mciteBstWouldAddEndPuncttrue
\mciteSetBstMidEndSepPunct{\mcitedefaultmidpunct}
{\mcitedefaultendpunct}{\mcitedefaultseppunct}\relax
\EndOfBibitem
\bibitem[Curtright and Zachos(2012)Curtright, and Zachos]{Curtright2011}
Curtright,~T.~L.; Zachos,~C.~K. Quantum Mechanics in Phase Space. \emph{Asia
  Pacific Physics Newsletter} \textbf{2012}, \emph{01}, 37\relax
\mciteBstWouldAddEndPuncttrue
\mciteSetBstMidEndSepPunct{\mcitedefaultmidpunct}
{\mcitedefaultendpunct}{\mcitedefaultseppunct}\relax
\EndOfBibitem
\bibitem[Curtright \latin{et~al.}(2013)Curtright, Fairlie, and
  Zachos]{Curtright2013}
Curtright,~T.; Fairlie,~D.~B.; Zachos,~C.~K. \emph{A Concise Treatise on
  Quantum Mechanics in Phase Space}; World Scientific, 2013\relax
\mciteBstWouldAddEndPuncttrue
\mciteSetBstMidEndSepPunct{\mcitedefaultmidpunct}
{\mcitedefaultendpunct}{\mcitedefaultseppunct}\relax
\EndOfBibitem
\bibitem[Gardiner(1985)]{Gardiner1985}
Gardiner,~C. \emph{Stochastic methods}; Springer, 1985\relax
\mciteBstWouldAddEndPuncttrue
\mciteSetBstMidEndSepPunct{\mcitedefaultmidpunct}
{\mcitedefaultendpunct}{\mcitedefaultseppunct}\relax
\EndOfBibitem
\bibitem[Zurek(2003)]{Zurek2003}
Zurek,~W.~H. Decoherence, einselection, and the quantum origins of the
  classical. \emph{Rev. Mod. Phys.} \textbf{2003}, \emph{75}, 715--775\relax
\mciteBstWouldAddEndPuncttrue
\mciteSetBstMidEndSepPunct{\mcitedefaultmidpunct}
{\mcitedefaultendpunct}{\mcitedefaultseppunct}\relax
\EndOfBibitem
\bibitem[Cabrera \latin{et~al.}(2015)Cabrera, Bondar, Jacobs, and
  Rabitz]{Cabrera2015}
Cabrera,~R.; Bondar,~D.~I.; Jacobs,~K.; Rabitz,~H.~A. Efficient method to
  generate time evolution of the Wigner function for open quantum systems.
  \emph{Phys. Rev. A} \textbf{2015}, \emph{92}, 042122\relax
\mciteBstWouldAddEndPuncttrue
\mciteSetBstMidEndSepPunct{\mcitedefaultmidpunct}
{\mcitedefaultendpunct}{\mcitedefaultseppunct}\relax
\EndOfBibitem
\bibitem[Berman \latin{et~al.}(1992)Berman, Kosloff, and Tal-Ezer]{Berman1992}
Berman,~M.; Kosloff,~R.; Tal-Ezer,~H. Solution of the time-dependent
  Liouville-von Neumann equation: dissipative evolution. \emph{J. Phys. A}
  \textbf{1992}, \emph{25}, 1283\relax
\mciteBstWouldAddEndPuncttrue
\mciteSetBstMidEndSepPunct{\mcitedefaultmidpunct}
{\mcitedefaultendpunct}{\mcitedefaultseppunct}\relax
\EndOfBibitem
\bibitem[Gao(1997)]{Gao1997}
Gao,~S. Dissipative Quantum Dynamics with a Lindblad Functional. \emph{Phys.
  Rev. Lett.} \textbf{1997}, \emph{79}, 3101--3104\relax
\mciteBstWouldAddEndPuncttrue
\mciteSetBstMidEndSepPunct{\mcitedefaultmidpunct}
{\mcitedefaultendpunct}{\mcitedefaultseppunct}\relax
\EndOfBibitem
\bibitem[Vacchini and Hornberger(2009)Vacchini, and Hornberger]{Vacchini2009}
Vacchini,~B.; Hornberger,~K. Quantum linear Boltzmann equation. \emph{Phys.
  Rep.} \textbf{2009}, \emph{478}, 71--120\relax
\mciteBstWouldAddEndPuncttrue
\mciteSetBstMidEndSepPunct{\mcitedefaultmidpunct}
{\mcitedefaultendpunct}{\mcitedefaultseppunct}\relax
\EndOfBibitem
\bibitem[Caldeira and Leggett(1983)Caldeira, and Leggett]{Caldeira1983}
Caldeira,~A.~O.; Leggett,~A.~J. {Path integral approach to quantum Brownian
  motion}. \emph{Physica A} \textbf{1983}, \emph{121}, 587--616\relax
\mciteBstWouldAddEndPuncttrue
\mciteSetBstMidEndSepPunct{\mcitedefaultmidpunct}
{\mcitedefaultendpunct}{\mcitedefaultseppunct}\relax
\EndOfBibitem
\bibitem[Ford and Kac(1987)Ford, and Kac]{Ford1987}
Ford,~G.~W.; Kac,~M. On the quantum Langevin equation. \emph{J. Stat. Phys.}
  \textbf{1987}, \emph{46}, 803--810\relax
\mciteBstWouldAddEndPuncttrue
\mciteSetBstMidEndSepPunct{\mcitedefaultmidpunct}
{\mcitedefaultendpunct}{\mcitedefaultseppunct}\relax
\EndOfBibitem
\bibitem[Ford \latin{et~al.}(1988)Ford, Lewis, and O'Connell]{Ford1988}
Ford,~G.~W.; Lewis,~J.~T.; O'Connell,~R.~F. Quantum langevin equation.
  \emph{Phys. Rev. A} \textbf{1988}, \emph{37}, 4419\relax
\mciteBstWouldAddEndPuncttrue
\mciteSetBstMidEndSepPunct{\mcitedefaultmidpunct}
{\mcitedefaultendpunct}{\mcitedefaultseppunct}\relax
\EndOfBibitem
\bibitem[Hu \latin{et~al.}(1992)Hu, Paz, and Zhang]{Hu1992}
Hu,~B.~L.; Paz,~J.~P.; Zhang,~Y. Quantum Brownian motion in a general
  environment: Exact master equation with nonlocal dissipation and colored
  noise. \emph{Phys. Rev. D} \textbf{1992}, \emph{45}, 2843--2861\relax
\mciteBstWouldAddEndPuncttrue
\mciteSetBstMidEndSepPunct{\mcitedefaultmidpunct}
{\mcitedefaultendpunct}{\mcitedefaultseppunct}\relax
\EndOfBibitem
\bibitem[Bolivar(2005)]{Bolivar2005}
Bolivar,~A.~O. Quantum Tunneling at Zero Temperature in the Strong Friction
  Regime. \emph{Phys. Rev. Lett.} \textbf{2005}, \emph{94}, 026807\relax
\mciteBstWouldAddEndPuncttrue
\mciteSetBstMidEndSepPunct{\mcitedefaultmidpunct}
{\mcitedefaultendpunct}{\mcitedefaultseppunct}\relax
\EndOfBibitem
\bibitem[Cleary \latin{et~al.}(2011)Cleary, Coffey, Dowling, Kalmykov, and
  Titov]{Cleary2011}
Cleary,~L.; Coffey,~W.~T.; Dowling,~W.~J.; Kalmykov,~Y.~P.; Titov,~S.~V. Phase
  space master equations for quantum Brownian motion in a periodic potential:
  comparison of various kinetic models. \emph{J. Phys. A} \textbf{2011},
  \emph{44}, 475001\relax
\mciteBstWouldAddEndPuncttrue
\mciteSetBstMidEndSepPunct{\mcitedefaultmidpunct}
{\mcitedefaultendpunct}{\mcitedefaultseppunct}\relax
\EndOfBibitem
\bibitem[Bolivar(2015)]{Bolivar2015}
Bolivar,~A.~O. Non-Markovian quantum Brownian motion: a non-Hamiltonian
  approach. \emph{arXiv:1503.07951} \textbf{2015}, \relax
\mciteBstWouldAddEndPunctfalse
\mciteSetBstMidEndSepPunct{\mcitedefaultmidpunct}
{}{\mcitedefaultseppunct}\relax
\EndOfBibitem
\bibitem[Diosi(1993)]{Diosi1993}
Diosi,~L. On High-Temperature Markovian Equation for Quantum Brownian Motion.
  \emph{EPL} \textbf{1993}, \emph{22}, 1\relax
\mciteBstWouldAddEndPuncttrue
\mciteSetBstMidEndSepPunct{\mcitedefaultmidpunct}
{\mcitedefaultendpunct}{\mcitedefaultseppunct}\relax
\EndOfBibitem
\bibitem[Vacchini(2000)]{Vacchini2000}
Vacchini,~B. Completely Positive Quantum Dissipation. \emph{Phys. Rev. Lett.}
  \textbf{2000}, \emph{84}, 1374--1377\relax
\mciteBstWouldAddEndPuncttrue
\mciteSetBstMidEndSepPunct{\mcitedefaultmidpunct}
{\mcitedefaultendpunct}{\mcitedefaultseppunct}\relax
\EndOfBibitem
\bibitem[Petruccione and Breuer(2002)Petruccione, and Breuer]{Petruccione2002}
Petruccione,~F.; Breuer,~H.-P. \emph{The theory of open quantum systems};
  Oxford Univ. Press, 2002\relax
\mciteBstWouldAddEndPuncttrue
\mciteSetBstMidEndSepPunct{\mcitedefaultmidpunct}
{\mcitedefaultendpunct}{\mcitedefaultseppunct}\relax
\EndOfBibitem
\bibitem[Munro and Gardiner(1996)Munro, and Gardiner]{Munro1996}
Munro,~W.~J.; Gardiner,~C.~W. Non-rotating-wave master equation. \emph{Phys.
  Rev. A} \textbf{1996}, \emph{53}, 2633--2640\relax
\mciteBstWouldAddEndPuncttrue
\mciteSetBstMidEndSepPunct{\mcitedefaultmidpunct}
{\mcitedefaultendpunct}{\mcitedefaultseppunct}\relax
\EndOfBibitem
\bibitem[Stenholm(1997)]{Stenholm1997}
Stenholm,~S. Quantum theory of linear friction. \emph{Brazilian J. Phys.}
  \textbf{1997}, \emph{27}, 214--237\relax
\mciteBstWouldAddEndPuncttrue
\mciteSetBstMidEndSepPunct{\mcitedefaultmidpunct}
{\mcitedefaultendpunct}{\mcitedefaultseppunct}\relax
\EndOfBibitem
\bibitem[Wiseman and Munro(1998)Wiseman, and Munro]{Wiseman1998}
Wiseman,~H.; Munro,~W. {Comment on "Dissipative Quantum Dynamics with a
  Lindblad Functional"}. \emph{Phys. Rev. Lett.} \textbf{1998}, \emph{80},
  5702--5702\relax
\mciteBstWouldAddEndPuncttrue
\mciteSetBstMidEndSepPunct{\mcitedefaultmidpunct}
{\mcitedefaultendpunct}{\mcitedefaultseppunct}\relax
\EndOfBibitem
\bibitem[Kohen \latin{et~al.}(1997)Kohen, Marston, and Tannor]{Kohen1997}
Kohen,~D.; Marston,~C.~C.; Tannor,~D.~J. Phase space approach to theories of
  quantum dissipation. \emph{J. Chem. Phys.} \textbf{1997}, \emph{107},
  5236--5253\relax
\mciteBstWouldAddEndPuncttrue
\mciteSetBstMidEndSepPunct{\mcitedefaultmidpunct}
{\mcitedefaultendpunct}{\mcitedefaultseppunct}\relax
\EndOfBibitem
\bibitem[Ford and O'Connell(2001)Ford, and O'Connell]{Ford2001}
Ford,~G.~W.; O'Connell,~R.~F. Exact solution of the Hu-Paz-Zhang master
  equation. \emph{Phys. Rev. D} \textbf{2001}, \emph{64}, 105020\relax
\mciteBstWouldAddEndPuncttrue
\mciteSetBstMidEndSepPunct{\mcitedefaultmidpunct}
{\mcitedefaultendpunct}{\mcitedefaultseppunct}\relax
\EndOfBibitem
\bibitem[Fleming \latin{et~al.}(2011)Fleming, Roura, and Hu]{Fleming2011}
Fleming,~C.~H.; Roura,~A.; Hu,~B.~L. Exact analytical solutions to the master
  equation of quantum Brownian motion for a general environment. \emph{Ann.
  Phys.} \textbf{2011}, \emph{326}, 1207--1258\relax
\mciteBstWouldAddEndPuncttrue
\mciteSetBstMidEndSepPunct{\mcitedefaultmidpunct}
{\mcitedefaultendpunct}{\mcitedefaultseppunct}\relax
\EndOfBibitem
\bibitem[Ford and O'Connell(2005)Ford, and O'Connell]{Ford2005}
Ford,~G.~W.; O'Connell,~R.~F. Limitations on the utility of exact master
  equations. \emph{Ann. Phys.} \textbf{2005}, \emph{319}, 348--363\relax
\mciteBstWouldAddEndPuncttrue
\mciteSetBstMidEndSepPunct{\mcitedefaultmidpunct}
{\mcitedefaultendpunct}{\mcitedefaultseppunct}\relax
\EndOfBibitem
\bibitem[Bondar \latin{et~al.}(2012)Bondar, Cabrera, Lompay, Ivanov, and
  Rabitz]{Bondar2011c}
Bondar,~D.~I.; Cabrera,~R.; Lompay,~R.~R.; Ivanov,~M.~Y.; Rabitz,~H.~A.
  {Operational Dynamic Modeling Transcending Quantum and Classical Mechanics}.
  \emph{Phys. Rev. Lett.} \textbf{2012}, \emph{109}, 190403\relax
\mciteBstWouldAddEndPuncttrue
\mciteSetBstMidEndSepPunct{\mcitedefaultmidpunct}
{\mcitedefaultendpunct}{\mcitedefaultseppunct}\relax
\EndOfBibitem
\bibitem[Bondar \latin{et~al.}(2013)Bondar, Cabrera, Zhdanov, and
  Rabitz]{Bondar2013a}
Bondar,~D.~I.; Cabrera,~R.; Zhdanov,~D.~V.; Rabitz,~H.~A. {Wigner phase-space
  distribution as a wave function}. \emph{Phys. Rev. A} \textbf{2013},
  \emph{88}, 052108\relax
\mciteBstWouldAddEndPuncttrue
\mciteSetBstMidEndSepPunct{\mcitedefaultmidpunct}
{\mcitedefaultendpunct}{\mcitedefaultseppunct}\relax
\EndOfBibitem
\bibitem[Bondar \latin{et~al.}(2013)Bondar, Cabrera, and Rabitz]{Bondar2011}
Bondar,~D.~I.; Cabrera,~R.; Rabitz,~H.~A. {Conceptual inconsistencies in
  finite-dimensional quantum and classical mechanics}. \emph{Phys. Rev. A}
  \textbf{2013}, \emph{88}, 012116\relax
\mciteBstWouldAddEndPuncttrue
\mciteSetBstMidEndSepPunct{\mcitedefaultmidpunct}
{\mcitedefaultendpunct}{\mcitedefaultseppunct}\relax
\EndOfBibitem
\bibitem[Zhdanov and Seideman(2015)Zhdanov, and Seideman]{Zhdanov2015}
Zhdanov,~D.~V.; Seideman,~T. Wigner representation of the rotational dynamics
  of rigid tops. \emph{Phys. Rev. A} \textbf{2015}, \emph{92}, 012129\relax
\mciteBstWouldAddEndPuncttrue
\mciteSetBstMidEndSepPunct{\mcitedefaultmidpunct}
{\mcitedefaultendpunct}{\mcitedefaultseppunct}\relax
\EndOfBibitem
\bibitem[Radonji\ifmmode~\acute{c}\else \'{c}\fi{}
  \latin{et~al.}(2014)Radonji\ifmmode~\acute{c}\else \'{c}\fi{},
  Popovi\ifmmode~\acute{c}\else \'{c}\fi{}, Prvanovi\ifmmode~\acute{c}\else
  \'{c}\fi{}, and Buri\ifmmode~\acute{c}\else \'{c}\fi{}]{Radonjic2014}
Radonji\ifmmode~\acute{c}\else \'{c}\fi{},~M.; Popovi\ifmmode~\acute{c}\else
  \'{c}\fi{},~D.~B.; Prvanovi\ifmmode~\acute{c}\else \'{c}\fi{},~S.;
  Buri\ifmmode~\acute{c}\else \'{c}\fi{},~N. Ehrenfest principle and unitary
  dynamics of quantum-classical systems with general potential interaction.
  \emph{Phys. Rev. A} \textbf{2014}, \emph{89}, 024104\relax
\mciteBstWouldAddEndPuncttrue
\mciteSetBstMidEndSepPunct{\mcitedefaultmidpunct}
{\mcitedefaultendpunct}{\mcitedefaultseppunct}\relax
\EndOfBibitem
\bibitem[Cabrera \latin{et~al.}(2014)Cabrera, Bondar, Campos, and
  Rabitz]{Cabrera2014}
Cabrera,~R.; Bondar,~D.~I.; Campos,~A.~G.; Rabitz,~H.~A. {Relativistic
  decoherence: Efficient numerical simulations of spin 1/2 relativistic open
  quantum systems}. \emph{arXiv:1409.1247} \textbf{2014}, \relax
\mciteBstWouldAddEndPunctfalse
\mciteSetBstMidEndSepPunct{\mcitedefaultmidpunct}
{}{\mcitedefaultseppunct}\relax
\EndOfBibitem
\bibitem[Lindblad(1976)]{Lindblad1976a}
Lindblad,~G. Brownian motion of a quantum harmonic oscillator. \emph{Rep. Math.
  Phys.} \textbf{1976}, \emph{10}, 393\relax
\mciteBstWouldAddEndPuncttrue
\mciteSetBstMidEndSepPunct{\mcitedefaultmidpunct}
{\mcitedefaultendpunct}{\mcitedefaultseppunct}\relax
\EndOfBibitem
\bibitem[Ferialdi(2016)]{Ferialdi2016}
Ferialdi,~L. Exact Closed Master Equation for Gaussian Non-Markovian Dynamics.
  \emph{Phys. Rev. Lett.} \textbf{2016}, \emph{116}, 120402\relax
\mciteBstWouldAddEndPuncttrue
\mciteSetBstMidEndSepPunct{\mcitedefaultmidpunct}
{\mcitedefaultendpunct}{\mcitedefaultseppunct}\relax
\EndOfBibitem
\bibitem[Mukamel(1999)]{Mukamel1999}
Mukamel,~S. \emph{Principles of Nonlinear Optical Spectroscopy}; Oxford
  University Press, 1999\relax
\mciteBstWouldAddEndPuncttrue
\mciteSetBstMidEndSepPunct{\mcitedefaultmidpunct}
{\mcitedefaultendpunct}{\mcitedefaultseppunct}\relax
\EndOfBibitem
\bibitem[Breuer(2007)]{Breuer2007}
Breuer,~H.-P. Non-{Markovian} generalization of the {Lindblad} theory of open
  quantum systems. \emph{Phys. Rev. A} \textbf{2007}, \emph{75}, 022103\relax
\mciteBstWouldAddEndPuncttrue
\mciteSetBstMidEndSepPunct{\mcitedefaultmidpunct}
{\mcitedefaultendpunct}{\mcitedefaultseppunct}\relax
\EndOfBibitem
\end{mcitethebibliography}
\end{document}